\documentclass[aps,prd,twocolumn,preprintnumbers,showpacs,nofootinbib,amssymb]{revtex4-1}

\usepackage{amsfonts,amsmath,units,wasysym,epsfig,graphicx,verbatim,color,subfigure,graphicx}
\usepackage{amsmath}
\usepackage{amssymb}
\usepackage{amsfonts}
\usepackage{bm}
\usepackage{color}

\def\be{\begin{equation}}
\def\ee{\end{equation}}
\def\bea{\begin{eqnarray}}
\def\eea{\end{eqnarray}}

\begin{document}


\newcommand{\rhat}{\hat{r}}
\newcommand{\iotahat}{\hat{\iota}}
\newcommand{\phihat}{\hat{\phi}}
\newcommand{\h}{\mathfrak{h}}
\newcommand{\IUCAA}{Inter-University Centre for Astronomy and Astrophysics, Post Bag 4, Ganeshkhind, Pune 411 007, India}
\newcommand{\ICTS}{International Centre for Theoretical Sciences, Tata Institute of Fundamental Research,
Bangalore 560012, India}
\newcommand{\WSU}{Department of Physics \& Astronomy, Washington State University,
1245 Webster, Pullman, WA 99164-2814, U.S.A \\}
\newcommand{\GU}{Department of Physics, Gujarat University, 120 Circular Road, Navrangpura, Ahmedabad, Gujarat 380009, India}

\title{Observational constraints on spinning, relativistic Bose-Einstein condensate stars}

\author{Arunava Mukherjee}
\email{arunava.mukherjee@icts.res.in}
\affiliation{\IUCAA}
\affiliation{\ICTS}

\author{Shreya Shah}
\email{Deceased 14 June 2014}
\affiliation{\GU}

\author{Sukanta Bose}
\email{sukanta@wsu.edu}
\affiliation{\IUCAA}
\affiliation{\WSU}

\date{\today}

\pacs{95.85.Sz, 04.30.Db, 97.60.Jd}

\begin{abstract}

Bose-Einstein condensates (BECs) have been proposed as candidate states of matter for the interior of neutron stars. Specifically, Chavanis and Harko obtained the mass-radius relation for a BEC star and proposed that the recently discovered neutron stars with masses around 2$M_\odot$ are BEC stars. 
They employed a barotropic equation of state (EOS), with one free parameter, that was first found by Colpi, Wasserman, and Shapiro (CSW), to describe them and 
derive stable equilibrium configurations of spinning BEC stars in General Relativity. 
In this work we show that while it is true that BECs allow for compact object masses 
as heavy as the heaviest observed ones, such stars cannot simultaneously have radii that are small enough to be consistent with the latest observations, in spite of the flexibility available in the EOS in the form of the free parameter.
In fact, our conclusion applies to any spinning relativistic boson star that obeys the CSW EOS.



\end{abstract}
\preprint{[LIGO-P1400179]}

\maketitle



\section{Introduction}

Neutron stars are natural laboratories for studying the extreme physical conditions that support the existence of ultra-dense degenerate nuclear matter, which  
is held together by its own strong gravity. Comprehending the structure and composition of these compact stars is challenging. Although, a complete understanding of the neutron star interior is yet to emerge, several equations of state have been proposed to describe it. The core of this compact star is made of hadronic matter and the interaction of such a many-body system is governed by the laws of quantum chromodynamics (QCD). 
Matter under such conditions is expected to be described by 
an experimentally unexplored and a completely different regime of the QCD phase diagram than the one, e.g., of hot dense matter such as the quark-gluon plasma. This is because the former has a very high chemical potential and supra-nuclear density, along with a very low temperature compared to the Fermi temperature, $T_F$. Deriving the properties of this phase from any first principle calculations, such as in lattice gauge theory, is currently not a viable proposition.

The strong self-gravitational field of a neutron star\footnote{Unless otherwise specified, in this paper, by a ``neutron star'' we mean any compact star that may be made of only neutron rich nuclear matter (truly neutron star), only quark matter at the core (quark star), a combination of both (hybrid star) or a star with a Bose-Einstein condensate (BEC) of hadronic or quark matter.} harbors extreme physical conditions that can create a unique fundamental phase of hadronic matter. This dense degenerate nuclear matter cannot be created in any terrestrial laboratory, e.g., RHIC, CERN, or any other particle collider where the center-of-mass energy is significantly higher than $T_F$ at a comparable density. 

It has been speculated for a long time that the condition in the neutron star interior may be suitable enough to support Bose-Einstein condensates of different types of nuclear matter, such as pions, kaons or even neutrons, without necessarily involving any exotic particles, e.g., certain flavors of quarks. Sawyer \cite{Sawyer:1972cq}, Barshay et al. \cite{Barshay:1973ju} and Baym \cite{Baym:1973zk} were among the first to propose that super-dense nuclear matter in the interior of a neutron star can undergo phase transition to produce $\pi^{-}$ condensation. Glendenning et al. \cite{Glendenning:1982ca, Glendenning:1982nn} formulated the theoretical background of isospin asymmetric nuclear
matter with both normal and pion condensed states 
in neutron stars and derived a few 
candidate equations of state, some of which can produce a $\sim 2$ M$_{\odot}$ star. 

%
%

Recently a couple of massive neutron stars, with masses of $2.01 \pm 0.04~M_\odot$ and $1.97 \pm 0.04~M_\odot$, were discovered by Antoniadis et al. \cite{Antoniadis:2013pzd} and Demorest et al. \cite{Demorest:2010bx}, respectively, which triggered multiple papers proposing various models of cold, dense, degenerate nuclear matter that can support stable configurations of self-gravitating fluid with those observed large mass values \cite{Prakash:2014tva,Pagliara:2014gja,Lattimer:2013zqa,Yasutake:2014oxa,Fantina:2014qea}. 
Specifically, superconductivity and superfluidity have been hypothesized in the past
to exist in neutron stars to explain certain observational effects (see, e.g., Ref.~\cite{Page:2000wt}). Neutrons bound in Cooper pairs can form a BEC. Each pair can be treated as a composite boson with an effective mass nearly equal to twice the mass of a neutron, or less, 
depending on how large the binding energy is. A microscopically exact way of treating such a system is provided by the theory of BCS-BEC crossover, which describes a transition from the quantum state of superfluidity (BCS phase) to a BEC. Nishida and Abuki \cite{Nishida:2005ds} explained the basic concept of this crossover and showed that these two states are smoothly connected without a phase transition. 

Just as with masses, several observational measurements exist of the radii of neutron stars \cite{Lattimer:2013hma,Guillot:2013wu,Ozel:2011ht,Guver:2010td,Ozel:2010fw}. Although simultaneous measurements of both the mass and the radius of the same object are available for a few cases \cite{Ozel:2011ht,Guver:2010td,Ozel:2010fw}, more reliable and tighter constraints come from a collection of different methods applied to various sources. This makes it possible to seek consistency among all those astronomical observations for any particular equation of state. What has been possible is to obtain constraints on it by examining the distributions of the measured masses and radii of various sources (see, e.g., the review \cite{Lattimer:2012nd} and the references therein). Therefore, whenever new observations are made, such as with the aforementioned two heaviest neutron stars, it is important to revisit how those constraints get revised and what equations of state get ruled out or in. Note that no radius measurements are available for these (heaviest) neutron stars till date.

In this paper, we study the structural properties of neutron stars, particularly, mass ($M$), radius ($R$) and spin, formed from a BEC of neutron Cooper pairs, the equation of state (EOS) for which was first introduced by Colpi, Shapiro, Wasserman (CSW) \cite{Colpi:1986ye}. Chavanis and Harko \cite{Cha} hypothesized that the recently observed massive neutron stars~\cite{Antoniadis:2013pzd,Demorest:2010bx} are BEC stars obeying the CSW EOS. This proposal was based on their demonstration that the CSW EOS, when applied to the Tolman-Oppenheimer-Volkoff (TOV) equations of relativistic stellar structure, allows for the observed large neutron star masses. This in turn is facilitated by a range of values that the scattering length and the mass of the Cooper pair of neutrons can assume, both of which are free parameters in their model for the EOS. 
In the literature, however, the CSW EOS has been adopted for studying not just the possibility of BECs as the constituents of stellar mass compact objects but also the viability of equilibrium configurations of boson stars \cite{Jetzer:1991jr,Liddle:1993ha,Schunck:2003kk,Seidel:1990jh}. Indeed, boson stars have been considered as possible dark matter candidates \cite{Lee:1995af,Nucamendi:2000jw}. In the cosmological context, boson stars have been proposed to have a range of masses, including those comparable to supermassive black hole masses (see, e.g., Refs. \cite{Schunck:1997dn,Dabrowski:1998ac,Torres:2000dw}). Furthermore, various people have studied observational consequences of the existence of boson stars, such as through electromagnetic observations \cite{Guzman:2005bs} and gravitational waves (see, e.g., Refs. \cite{Liebling:2012fv,Yunes:2013dva,Yuan:2004sv} and the references therein). Therefore, if the aforementioned hypothesis of Chavanis and Harko is indeed true, then it can have important implications for these proposed observations.

With this in mind, here we revisit the theoretically allowed equilibrium configurations of boson stars, specifically, to test the predictions of Chavanis and Harko against astronomical observations of the masses and radii of neutron stars.
We build on their work by generalizing it to include spin, with different scattering lengths and Cooper-pair masses, to compute the observationally verifiable mass-radius relations. 
As shown in the subsequent sections, unfortunately for this model, we find that the radius predicted by the CSW EOS is too large to be consistent with astronomical observations.


\section{Computing stellar structure with CSW EOS} 

\label{effect_a_EOS}

In Ref. \cite{Cha} Chavanis and Harko proposed that in the neutron star interior a BEC can exist in the form of neutron Cooper-pairs. They argued that this BEC can be described by the CSW EOS:
\begin{equation}
\label{CSW_EOS}
P=\frac{c^4}{36K}\left\lbrack \left (1+\frac{12K}{c^2}\rho \right )^{1/2}-1\right\rbrack^2 \\,
\end{equation}
where $P$ is the pressure of the fluid, $\rho = \epsilon/c^2$, $\epsilon$ is the total energy-density of the fluid, $c$ is the speed of light in vacuum, and
\begin{equation}
\label{K}
K=\frac{\lambda\hbar^3}{4m^4 c}\,.
\end{equation}
Above, $m$ is the mass of the neutron Cooper-pair, $\hbar$ 
is the Planck constant, and $\lambda$ is a dimensionless quantity defined by
\begin{equation}
\label{lambda}
\lambda = (9.523 \times{8\pi}) \frac{a}{1\, {\rm fm}}\frac{m}{2 m_n}\,.
\end{equation}
Above, $a$ is the scattering length that defines a $\frac{1}{4}\lambda\phi^4$ interaction between condensed bosons and, via the relativistic Gross-Pitaeviskii equation, gives rise to the EOS defined in Eq. (\ref{CSW_EOS}). Note that $m$ is a free parameter. It obeys $m \leq 2 m_n$,  where $m_n$ is the neutron mass. $\lambda$ is the second free parameter in their model.




As is evident in Eq.~(\ref{K}), both $a$ and $m$ enter the EOS through the single parameter $K$. Therefore, a stable equilibrium stationary stellar configuration of a self-gravitating condensate corresponding to a specific value of $K$ can actually arise from multiple possible  values of $a$ and $m$.
Below, we use the values of stellar structural quantities, such as mass, radius, and spin, as obtained from astronomical observations of neutron stars to constrain the free parameter $K$ 
and, hence, the allowed values of both scattering length $a$ and Cooper-pair mass $m$.




In this paper we consider stationary axisymmetric equilibrium configurations of the star with four different stellar spin frequencies, as well as the static spherically symmetric non-spinning case, with perfect fluid matter.
We used the publicly available ``Rapidly Rotating Neutron Star'' (RNS) code \cite{Stergioulas:1994ea} (which follows the algorithm defined in Komatsu et al. \cite{Komatsu:1989zz} and Cook et al. \cite{Cook:1993qr}) 
to numerically obtain these configurations for stellar spin frequencies 250 Hz, 500 Hz, 750 Hz, and 1 kHz, as well as the non-spinning one for the CSW equation of state. 
(See Appendix \ref{app:RNS} for a brief description of the RNS code.)

For each value of the stellar spin frequency chosen here, we computed the gravitational mass $M$ (generally referred to as ``mass'' in this paper, unless mentioned otherwise) and the stellar radius $R$, both equatorial and polar, corresponding to a hundred different values of the central density of the star. The results are given in Sec.~\ref{effect_EOS_NS-structure}.




\subsection{The effect of the free parameter ($K$) in CSW EOS on the structure of a neutron star}
\label{effect_EOS_NS-structure}
\vspace*{-0.4cm}

The value of the EOS parameter $K$ neither is known from any calculations from first principles nor has been determined yet from any experimental data and, therefore, remains a free parameter.
Any change in $K$ will change the EOS (see Eqs.~(\ref{CSW_EOS})-(\ref{lambda})) and, thus, will also affect the observable structural quantities. Here, we discuss the effect of that parameter on the maximum gravitational mass of such a neutron star.

\begin{figure}[h]
\begin{center}
\vspace*{-0.4cm}
\begin{tabular}{c}
\hspace*{-1.25cm}
\includegraphics[width=0.6\textwidth,angle=0]{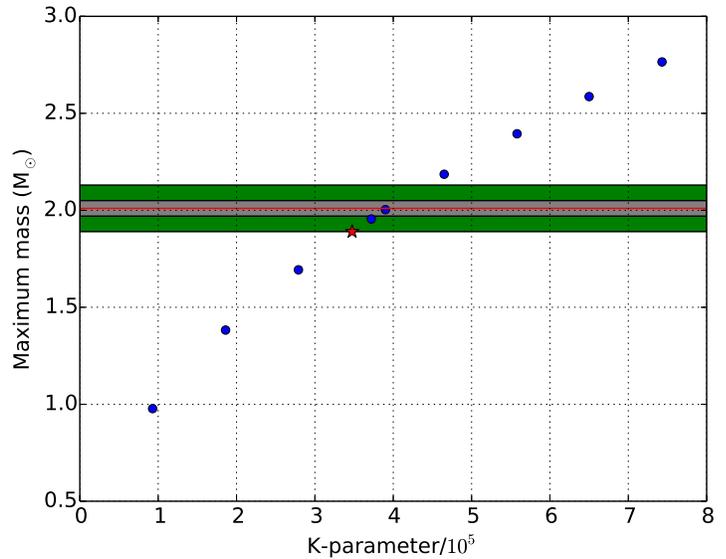} \\
\end{tabular}
\vspace*{-0.8cm}
\end{center}
\caption{The maximum gravitational mass of a BEC star obeying the CSW equation of state~\cite{Colpi:1986ye} is plotted as a function of the free parameter $K$ (in cm$^{5}$ g$^{-1}$s$^{-2}$) in (blue) dots and the (red) star. The thin horizontal (red) line shows the observed maximum mass of a neutron star and  the horizontal gray band shows the 1$\sigma$ error in that value~\cite{Antoniadis:2013pzd}. Here, the (red) star on the lower boundary of the green region is at $K =  3.475 \times 10^{5}$ cm$^{5}$ g$^{-1}$s$^{-2}$, and denotes the minimum value of $K$ allowed by observations when one widens the error on the observed maximum neutron star mass by $3\sigma$.
}
\label{K-parameter_max-mass-plot}
\end{figure}

\begin{figure}[h]
\begin{center}
\vspace*{-0.2cm}
\begin{tabular}{c}
\hspace*{-0.8cm}
\includegraphics[width=0.6\textwidth,angle=0]{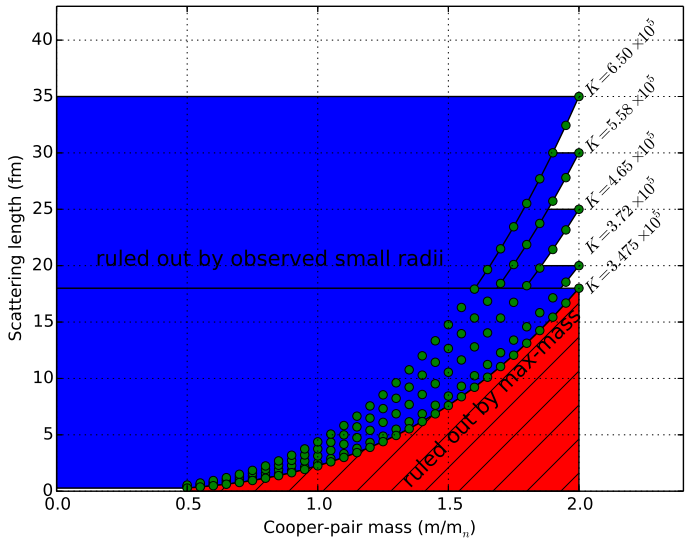} \\
\end{tabular}
\vspace*{-1.0cm}
\end{center}
\caption{The constraint on the scattering length $a$ and the Cooper-pair mass $m$ (in units of $m_n$) arising from the observational constraint on the free parameter $K$ (in cm$^{5}$ g$^{-1}$s$^{-2}$). Figure \ref{K-parameter_max-mass-plot} shows that the (red) hatched region, where $K < 3.475 \times 10^{5}$ cm$^{5}$ g$^{-1}$s$^{-2}$ is disallowed by the maximum observed mass of neutron stars. As argued in Sec. \ref{sec:astrobs}, the region (blue colored) where $K > 3.475 \times 10^{5}$ cm$^{5}$ g$^{-1}$s$^{-2}$ gets ruled out too by the fact that the observed neutron star radii are smaller than the values predicted by the CSW EOS.}
\label{Kam}
\end{figure}

In Ref.~\citep{Antoniadis:2013pzd} it was reported that a pulsar was observed with a gravitational mass as large as $2.01 \pm 0.04$ M$_\odot$. This is the most massive neutron star currently known for which such precise and reliable measurement was possible. As shown in Fig.~\ref{K-parameter_max-mass-plot}, we find that for $K < 3.475 \times 10^{5}$ cm$^{5}$ g$^{-1}$s$^{-2}$, the maximum neutron star mass $M_{\rm max}$  allowed by the CSW EOS is $1.89~M_\odot$, which is the maximum mass one finds for this pulsar after allowing for a $-3\sigma$ error in the observed value. Therefore, $K$ needs to be greater than $3.475 \times 10^{5}$ cm$^{5}$ g$^{-1}$s$^{-2}$ for the CSW EOS to allow for heavier maximum neutron star masses and, thereby, remain consistent with observations. Thus, in general all possible pairs of values of $a$ and $m$ that lie in the red hatched region in Fig.~\ref{Kam} rule out this kind of a BEC as the principal constituent of a neutron star. Hence, we compute the values of several stellar structure characteristics and plot them in Figs.~\ref{K-parameter_mass-radius_plot1} and~\ref{K-parameter_mass-radius_plot2} for $K \geq 3.475 \times 10^{5}$ cm$^{5}$ g$^{-1}$s$^{-2}$ in order to compare them with values obtained from astronomical observations. We summarize these observations in the next section before returning to study their implications on the results shown in Figs.~\ref{K-parameter_mass-radius_plot1} and~\ref{K-parameter_mass-radius_plot2}. 


\section{Constraints from astronomical observations}
\label{sec:astrobs}

Timing and spectral studies of several astrophysical phenomena in Galactic low-mass X-ray binary (LMXB) systems provide observational constraints on the mass-radius ($M$-$R$) relationship of neutron stars. Neutron stars in LMXB systems accrete matter from their companions. The accreted matter falls and accumulates on the neutron star surface. This accumulated matter, which is mostly hydrogen or helium, reaches the temperature, pressure and density conditions suitable for various thermonuclear reactions~\cite{Strohmayer:2003vf,StrohmayerBook:2006}.
These thermonuclear reactions in neutron stars in LMXB systems often occur intermittently and undergo run-away nuclear burning producing luminous flashes of X-ray radiation, generally termed as ``thermonuclear X-ray burst'' or ``type-I X-ray burst''. As a consequence, they are seen with several X-ray telescopes, particularly {\it Rossi Timing Explorer Satellite} ({\it RXTE}). 
In Ref.~\cite{Ozel:2010fw}, \"Ozel et al. studied three neutron star LMXBs, namely, 4U 1608--248, EXO 1745--248, and 4U 1820--30, where they analyzed photon energy spectra of a number of thermonuclear X-ray bursts from these three sources and used them to estimate the gravitational mass and radius of the respective neutron stars (see Fig.~1 in \cite{Ozel:2010fw}). The reported values for the 1-$\sigma$ and 2-$\sigma$ confidence contours imply that the masses and radii of these neutron stars lie in the range of $\approx 1.3-1.95~M_\odot$ and $\approx 8-12$ km, respectively. 

In another paper, \"Ozel et al. \cite{Ozel:2011ht} analyzed thermonuclear X-ray bursts from another neutron star LMXB system, viz., KS 1731--260, in the direction of the Galactic bulge. 
In this case, they were able to put an upper limit on the neutron star radius of $R \leq 12.5$ km at the 95\% confidence level, assuming its gravitational mass to be $M \leq 2.1~M_\odot$. 

Furthermore, G\"{u}ver et al.~\cite{Guver:2010td} studied the energy spectra in the X-ray band of {\it RXTE} observations of a number of thermonuclear bursts from the neutron star in the LMXB system 4U 1820--30 in the globular cluster NGC 6624, which provided for a well estimated distance to the source. They measured the mass and radius of the neutron star in that LMXB using the energy spectra extracted at the touchdown moments of several photospheric radius expansion (PRE) thermonuclear bursts 
and found them to be $M = 1.58 \pm 0.06~M_\odot$ and $R = 9.11 \pm 0.4$ km, respectively.



In Ref.~\cite{vanStraaten:2000gd}, van Straaten et al. detected a kHz QPO from a Galactic neutron star LMXB source 4U 0614+09 at $1329 \pm 4$ Hz. Although, the physical origin of these kHz QPOs is not unequivocally explained, the widely accepted view suggests that the centroid frequency corresponds to Keplerian orbital frequency around the neutron star. 
With that assumption the authors estimated the upper limits on both mass and radius of the neutron star to be $M \leq 1.9~M_\odot$ and $R \leq 15.2$ km, respectively.

Since 
X-ray observations are affected by systematic errors it is important to include different classes of sources in drawing inferences about neutron star radii.
Distinct from the aforementioned types of neutron star systems are 
quiescent low-mass X-ray binaries (qLMXBs) in globular clusters.
By analyzing the thermal spectra from a number of them, Guillot et al. \cite{Guillot:2013wu} reported that the inferred neutron star radii lie in the range $R = 7.6 - 10.4$~km (see Fig.~17 in Ref.~\cite{Guillot:2013wu}), which are again on the small side. Nonetheless, for a particular neutron star in $\omega$ Cen the same authors provided  an observational estimation of $R = 20 - 27$ km. This result, however, is beset by systematic errors. This is primarily due to the uncertainty in the amount of neutral hydrogen $n_{H}$ in the inter-stellar medium 
and the star's atmospheric chemical composition and is, therefore, highly debatable. To wit, 
by using different values of these parameters, Lattimer \& Steiner \cite{Lattimer:2013hma} reported a range of smaller values for the inferred radius of that neutron star in  $\omega$ Cen, namely, $R = 8-15$ km.

We now compare these astronomically observed values of neutron star radii with those that we obtain theoretically for the CSW EOS. The latter are 
presented in Fig.~\ref{K-parameter_mass-radius_plot1} \& Fig.~\ref{K-parameter_mass-radius_plot2}. 
We study the rotation induced changes in both the equatorial and the polar radii of spinning neutron stars (in some cases rapidly spinning) by computing the numerically exact solutions of stellar structure in full General Relativity. (Note that spin is expected to allow for a greater range of radii than the non-spinning configuration, thereby, improving the chances of overlap with observed radii.) Moreover, since most of the observed neutron stars are in LMXB systems, they are very likely to be spun up by accretion induced torque and may presently be spinning rapidly. Thus, the radius inferred observationally can vary from one method to another depending on whether it measures the equatorial or the polar value. 
This limitation notwithstanding, the measured radii must always lie between those two values.
In Figs.~\ref{K-parameter_mass-radius_plot1} and \ref{K-parameter_mass-radius_plot2} we present our theoretical results, which show that the smallest radius allowed by the CSW EOS, for any spin $\leq 1$ kHz and masses as high as $1.89~M_\odot$, is 17.5~km. The effect of rotation on the neutron star radius for the CSW EOS is not sufficiently large to explain the small observed values. This is the main conclusion of this work.

Limiting the spin frequency of the compact objects in our simulations to $\approx 1$ kHz is quite reasonable for these astrophysical neutron stars we observe in our Galaxy. The fastest-spinning pulsar known today is PSR J1748-2446ad, which has a spin frequency of 716 Hz. This corresponds to a spin-period of about 1.4ms. {\it RXTE} and {\it INTEGRAL} discovered a pulsar in 2007, namely, XTE J1739-28, which was initially claimed to be rotating at 1122 Hz \cite{Chakrabarty:2008gz}. But this result is not statistically significant. Nonetheless, we in fact considered stellar configurations with spins as high as 1kHz, which allows for smaller polar radii than non-spinning configurations. This gives the BEC models slightly more room for viability, vis a vis observational data. We could have considered somewhat higher spins too but as shown in Fig. 3, the reduced polar radii would still not make the BEC stars observationally consistent. Indeed, as shown in Fig. 3 the best BEC scenarios are for $K \lesssim 4 \times 10^{5} $ cm$^{5}$ g$^{-1}$s$^{-2}$, i.e., the two rows there. But even there, increasing the spin from 0 to 1kHz makes very little difference (i.e., less than a few percent) at the smallest radii observed in our configurations, which are $> 17.5$ km. Additionally, results from nuclear physics strongly constrain the radius of a neutron star to be $\lesssim 14$ km for masses $\gtrsim 1.2~M_\odot$ \cite{Lattimer:2012nd}.

Moreover, the distribution of all measured pulsar spin frequencies to date falls rather rapidly at the high-frequency end. In particular, the histogram of spin frequency distribution of the accreting neutron star in the LMXB systems, which are the major sources we considered here, show a significant cut-off at around 730 Hz \cite{Chakrabarty:2008gz}.

Therefore, the chance of having a neutron star above 1 kHz is extremely small. Moreover, we have considered a number of different LMXBs each having a different neutron star. Thus, it is highly unlikely that any neutron star will have a spin frequency higher than 1 kHz. If even a single neutron star in these systems has spin less than or around 1 kHz, it is sufficient to rule out the CSW EOS. In this way, the frequency band studied is representative of the spins as observed in nature.

Finally, a note on the turning points in each of the mass vs central density curves in Figs.~\ref{K-parameter_mass-radius_plot1} and \ref{K-parameter_mass-radius_plot2}: These points are associated with the maximum mass allowed by the corresponding configuration, and mark the onset of an instability in the fundamental radial pulsation mode of a non-spinning star \cite{Stergioulas:2003yp} and the secular axisymmetric instability in a spinning neutron star \cite{Friedman:1988er}. As was shown in Ref. \cite{Cook:1992myreference}, the onset of this instability separates stable configurations from ones that will collapse to form a black hole or some other compact object that is not a neutron condensate obeying the CSW EOS. The timescale of the development of this instability will be comparable to the viscous timescale of the EOS, but much smaller than the lifetime of a neutron star in an LMXB system. (Indeed, the largest glitch time ever observed in the large population of pulsars is also significantly smaller than the lifetime of the pulsars themselves \cite{Epstein:1992myreference1}.) Moreover, the collapse timescale will be comparable to the dynamical timescale of such a configuration and, thus, will be a negligible fraction of the lifetime of the neutron stars in the LMXBs considered here.


\section{Summary}

The masses of most components of double neutron star systems have been measured quite accurately but not their radius \cite{Heuvel:2007zp}. Moreover, the radii of the massive pulsars discovered in Refs. \cite{Antoniadis:2013pzd,Demorest:2010bx} are also unknown. It is worth noting here that the detected component(s) in these systems are all radio-pulsars. In the case of radio-pulsars any estimation of radius is practically impossible, which is unlike the case of accreting pulsars that are observable in X-rays. 

\begin{figure}[h]
\begin{center}
\vspace*{-0.3cm}
\begin{tabular}{lr}
\vspace*{-0.2cm}
\hspace*{-1.7cm}
\includegraphics[width=0.6\textwidth,angle=0]{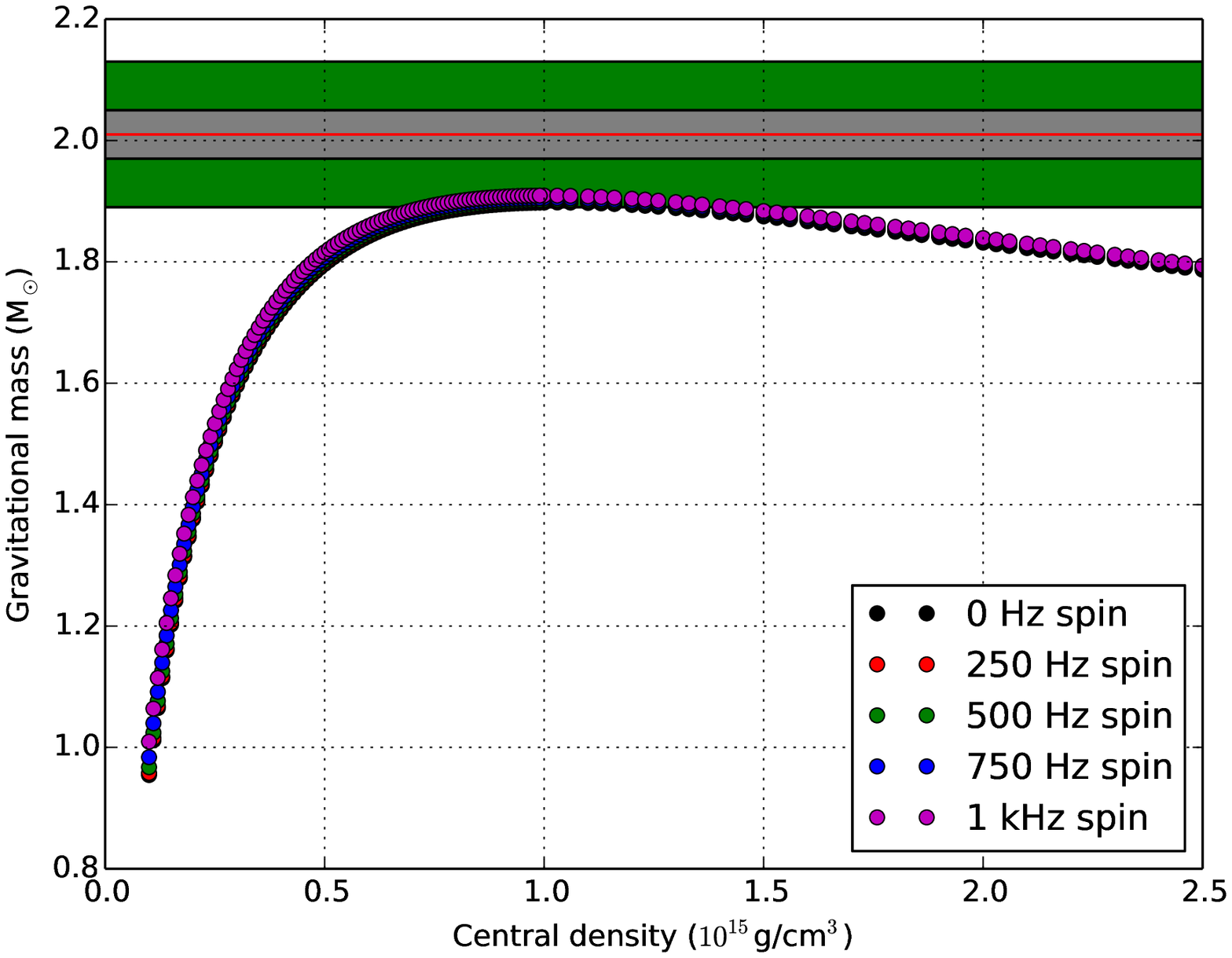} &
\hspace*{-1.1cm}
\includegraphics[width=0.6\textwidth,angle=0]{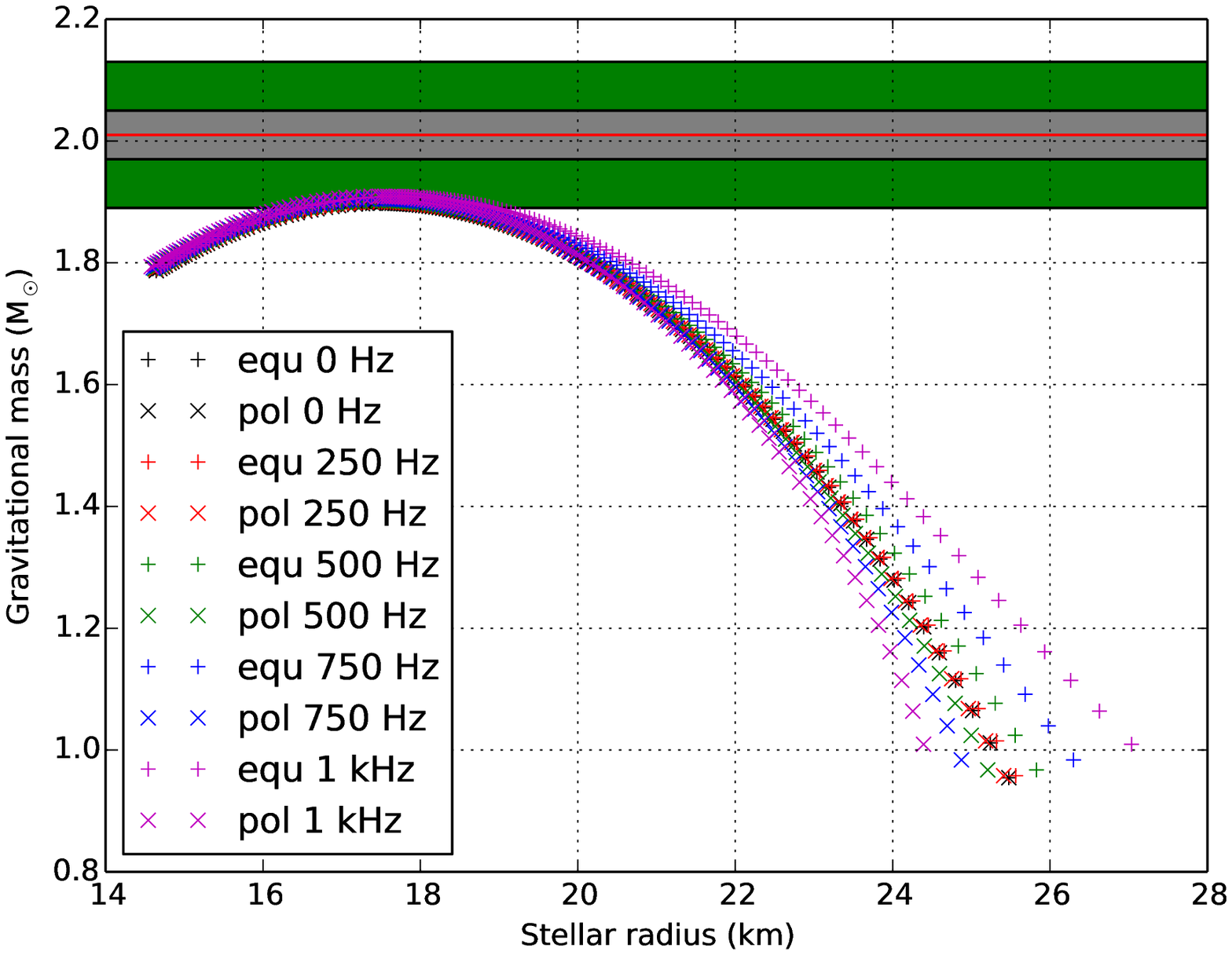} \\
\vspace*{-0.2cm}
\hspace*{-1.7cm}
\includegraphics[width=0.6\textwidth,angle=0]{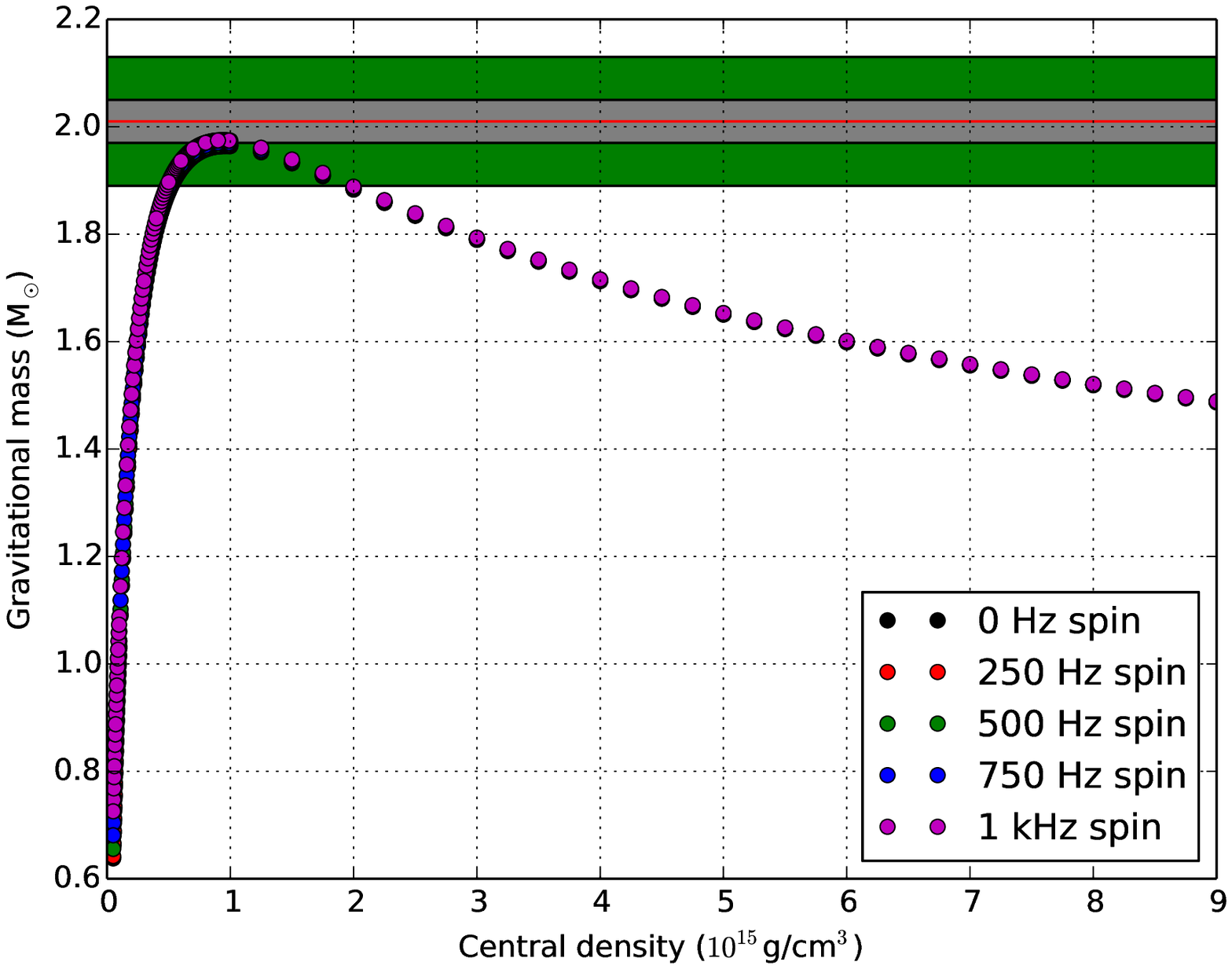} &
\hspace*{-1.1cm}
\includegraphics[width=0.6\textwidth,angle=0]{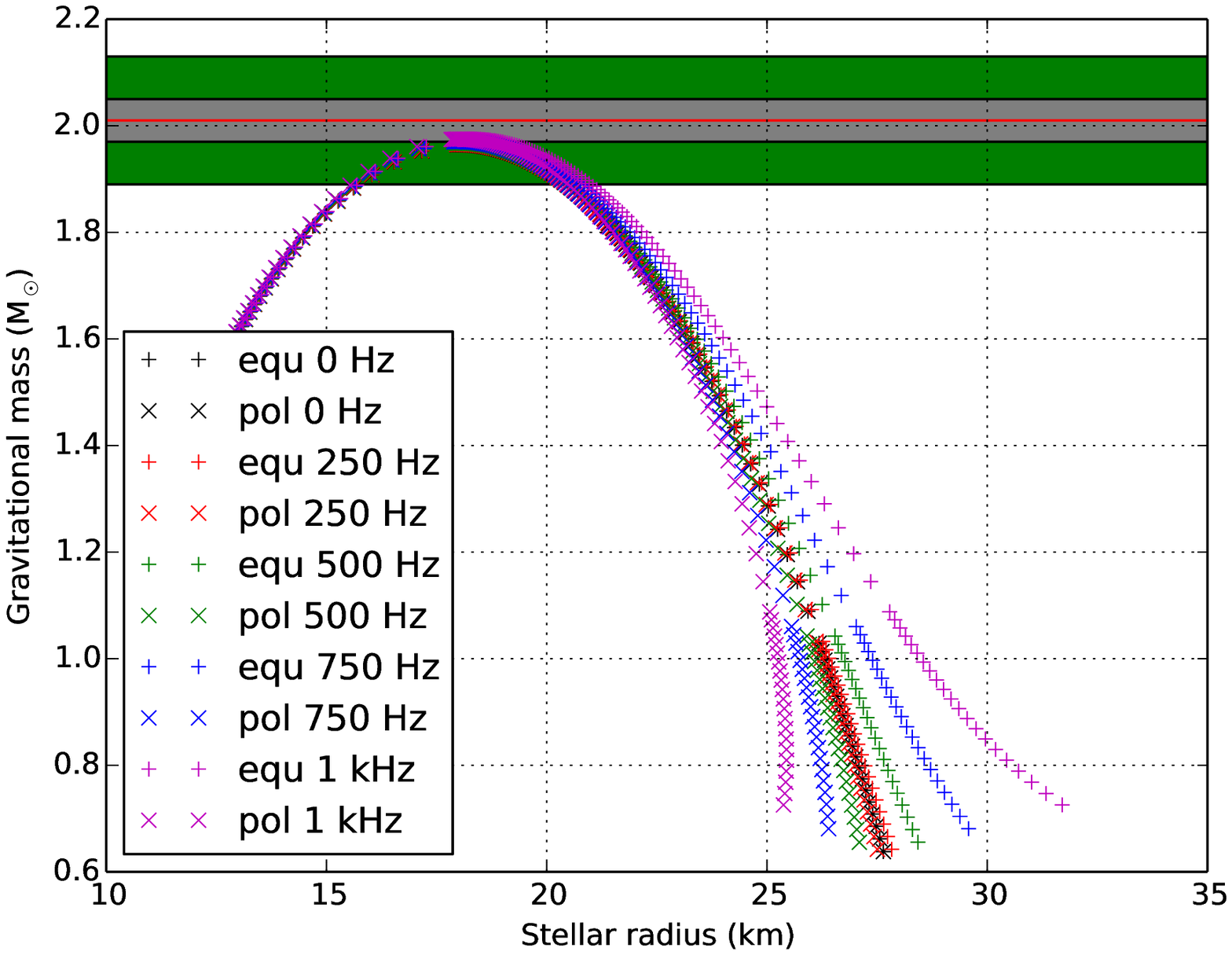} \\
\end{tabular}
\vspace*{-0.2cm}
\end{center}
\caption{The gravitational mass of neutron stars with a variety of spin frequencies and obeying the CSW EOS are plotted as a function of the central-density (left-column) and radius (right-column). Each row corresponds to a different value of the CSW EOS parameter (in cm$^{5}$ g$^{-1}$s$^{-2}$), namely, $K = 3.475 \times 10^{5}$ (top row) and $K=3.72 \times 10^{5}$ (bottom row). The equatorial (``equ'') and polar (``pol'') radii for a variety of neutron star spins along with the non-spinning (i.e., ``0 Hz spin'' case) are also shown here.}
\label{K-parameter_mass-radius_plot1}
\end{figure}

\begin{figure}[h]
\begin{center}
\vspace*{-0.3cm}
\begin{tabular}{lr}
\vspace*{-0.2cm}
\hspace*{-1.7cm}
\includegraphics[width=0.6\textwidth,angle=0]{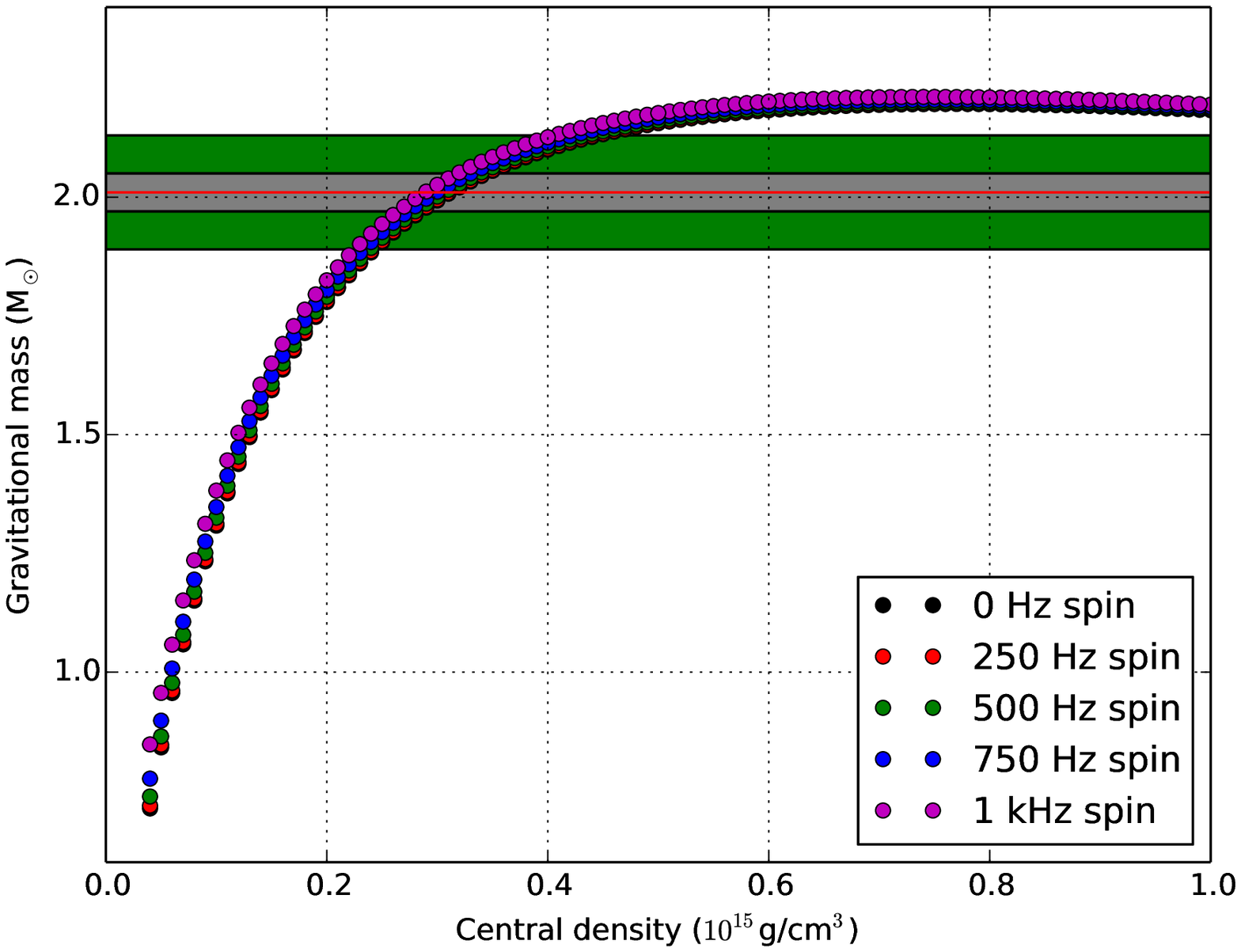} &
\hspace*{-1.1cm}
\includegraphics[width=0.6\textwidth,angle=0]{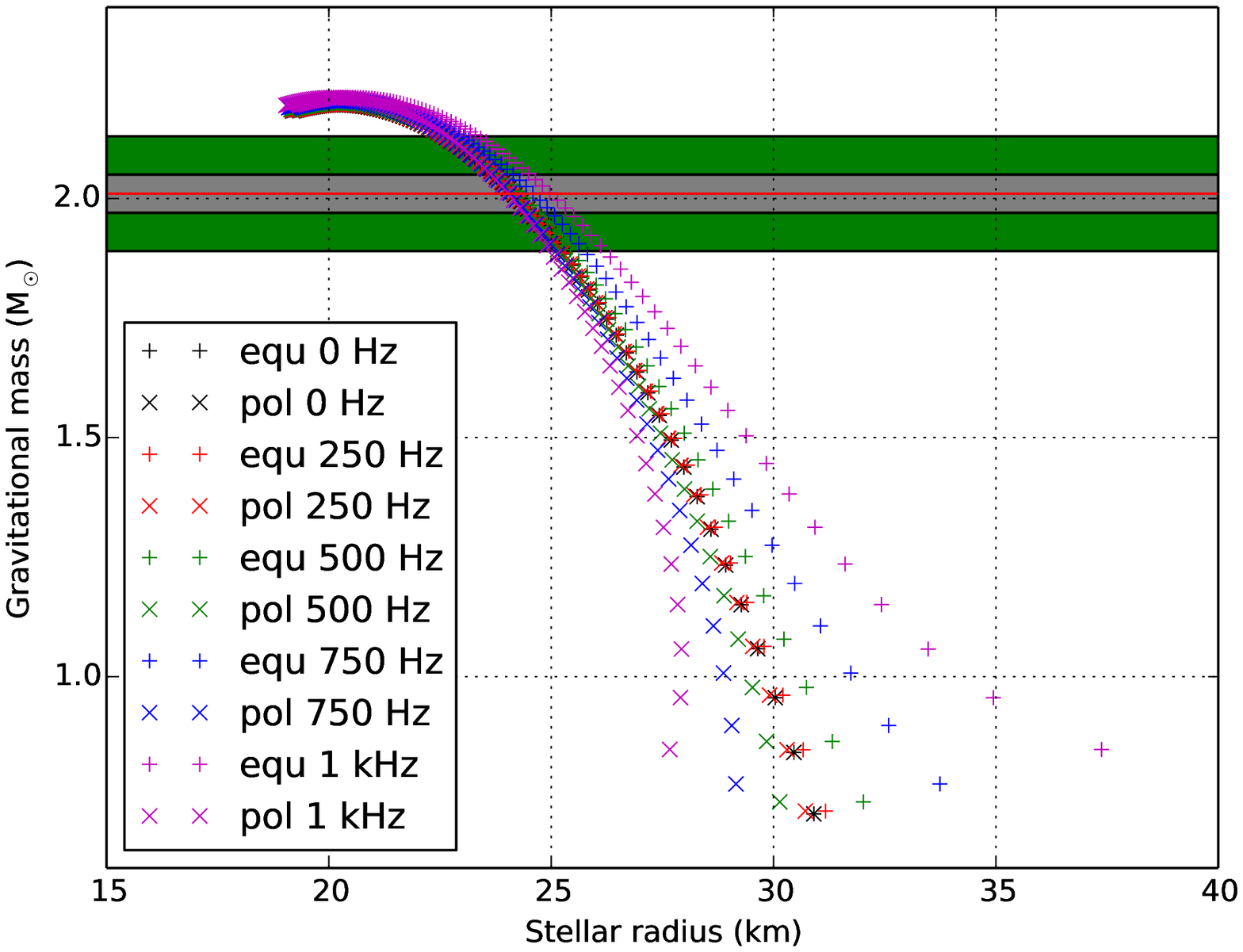} \\
\vspace*{-0.2cm}
\hspace*{-1.7cm}
\includegraphics[width=0.6\textwidth,angle=0]{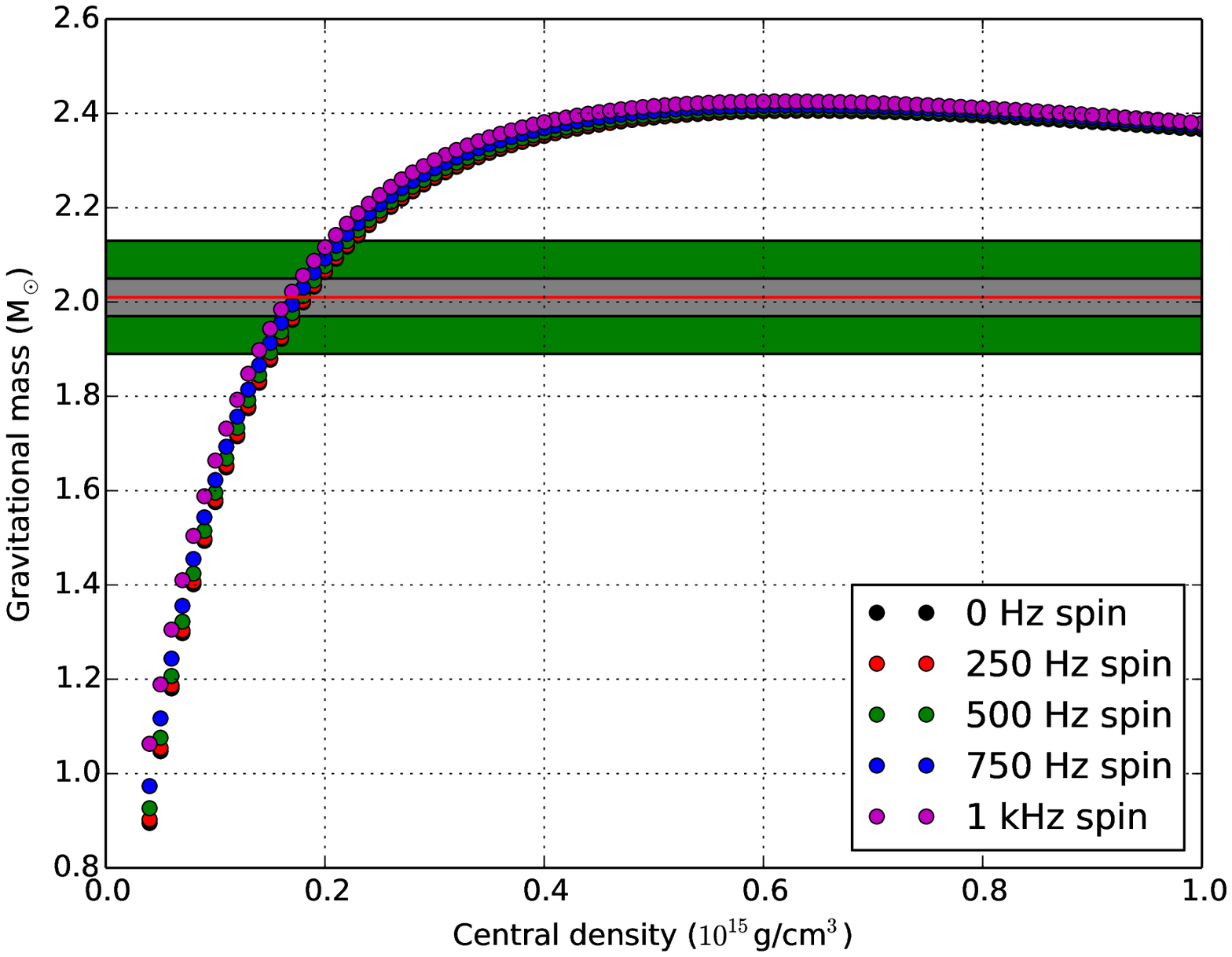} &
\hspace*{-1.1cm}
\includegraphics[width=0.6\textwidth,angle=0]{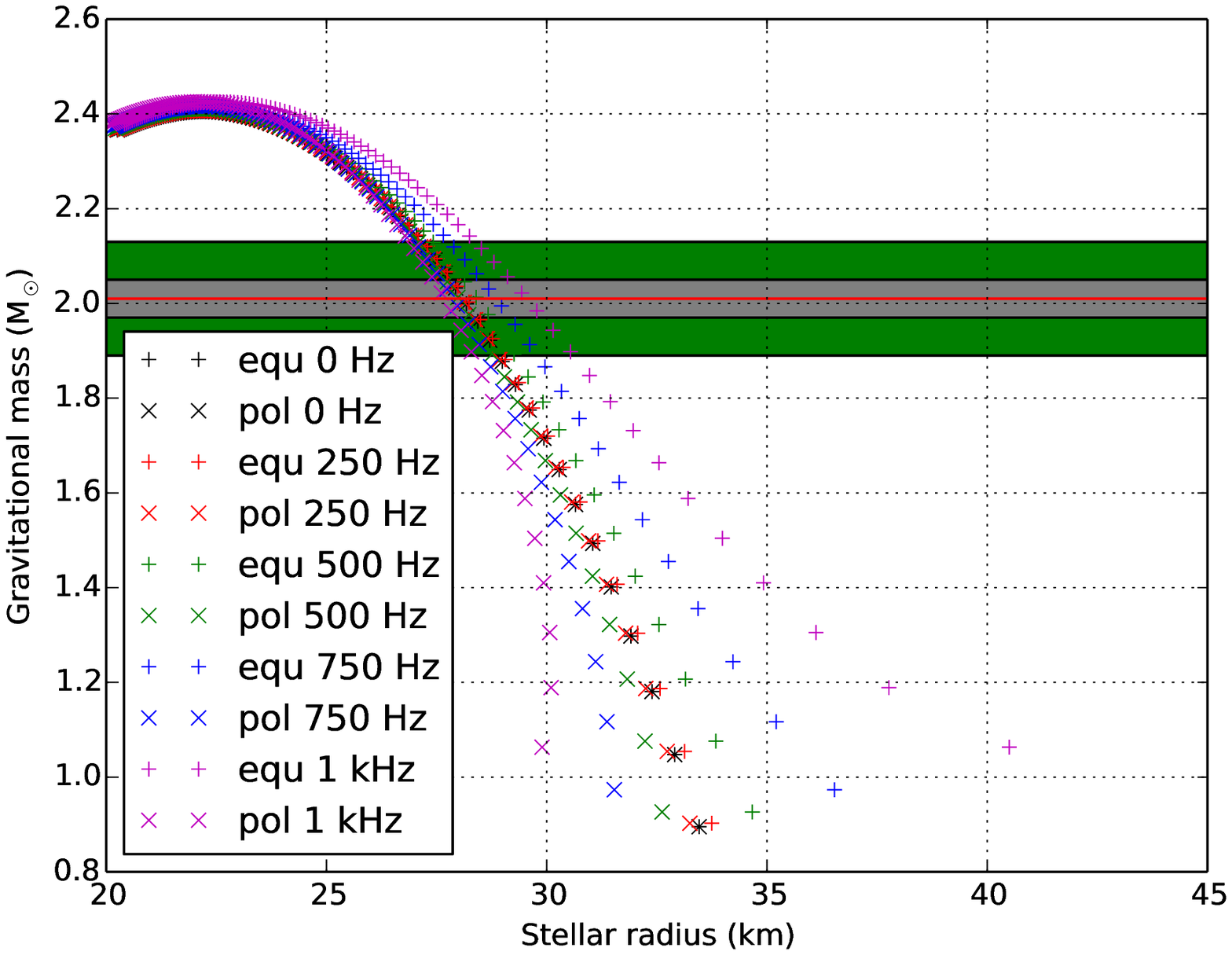} \\
\hspace*{-1.7cm}
\vspace*{-0.2cm}
\includegraphics[width=0.6\textwidth,angle=0]{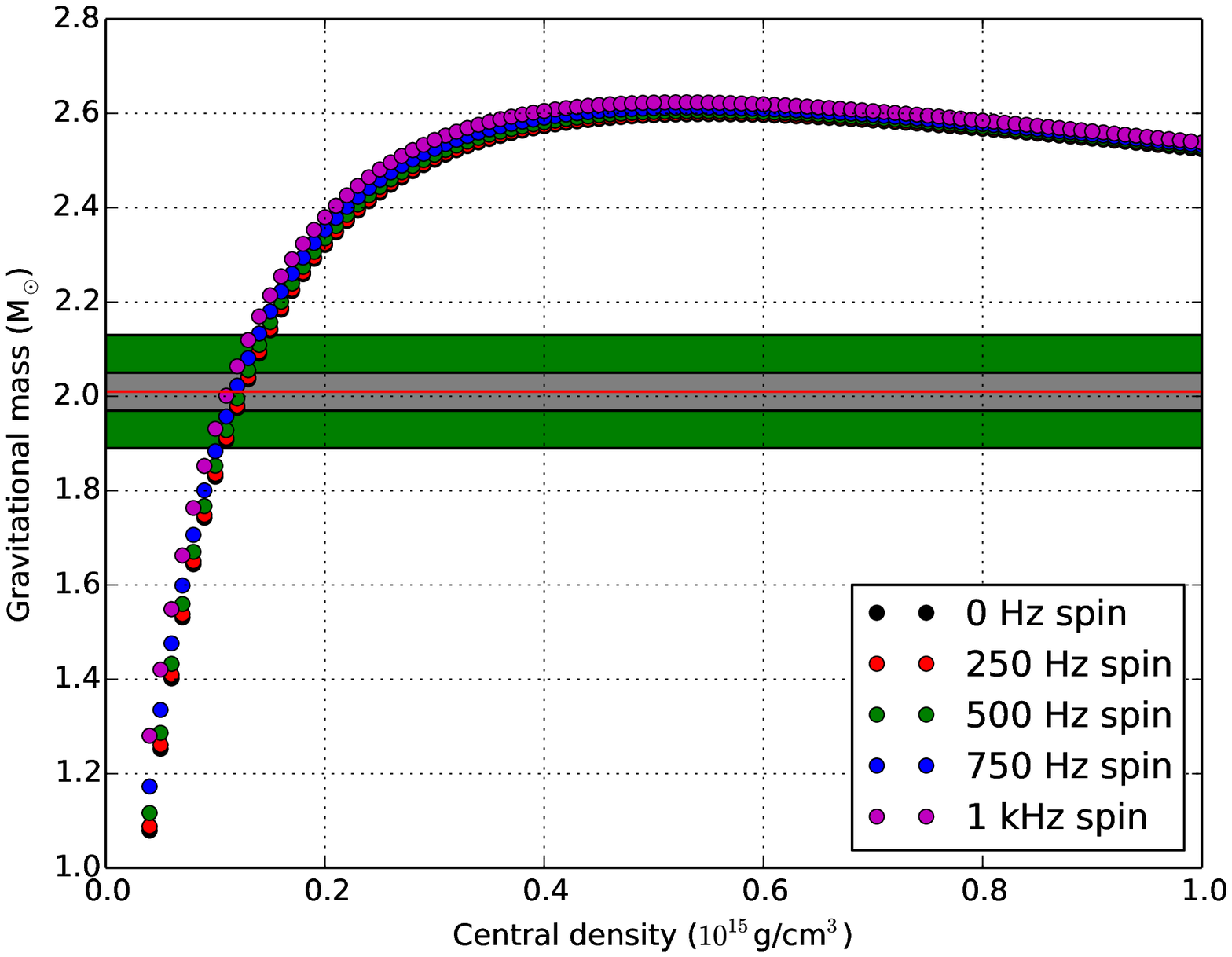} &
\hspace*{-1.1cm}
\includegraphics[width=0.6\textwidth,angle=0]{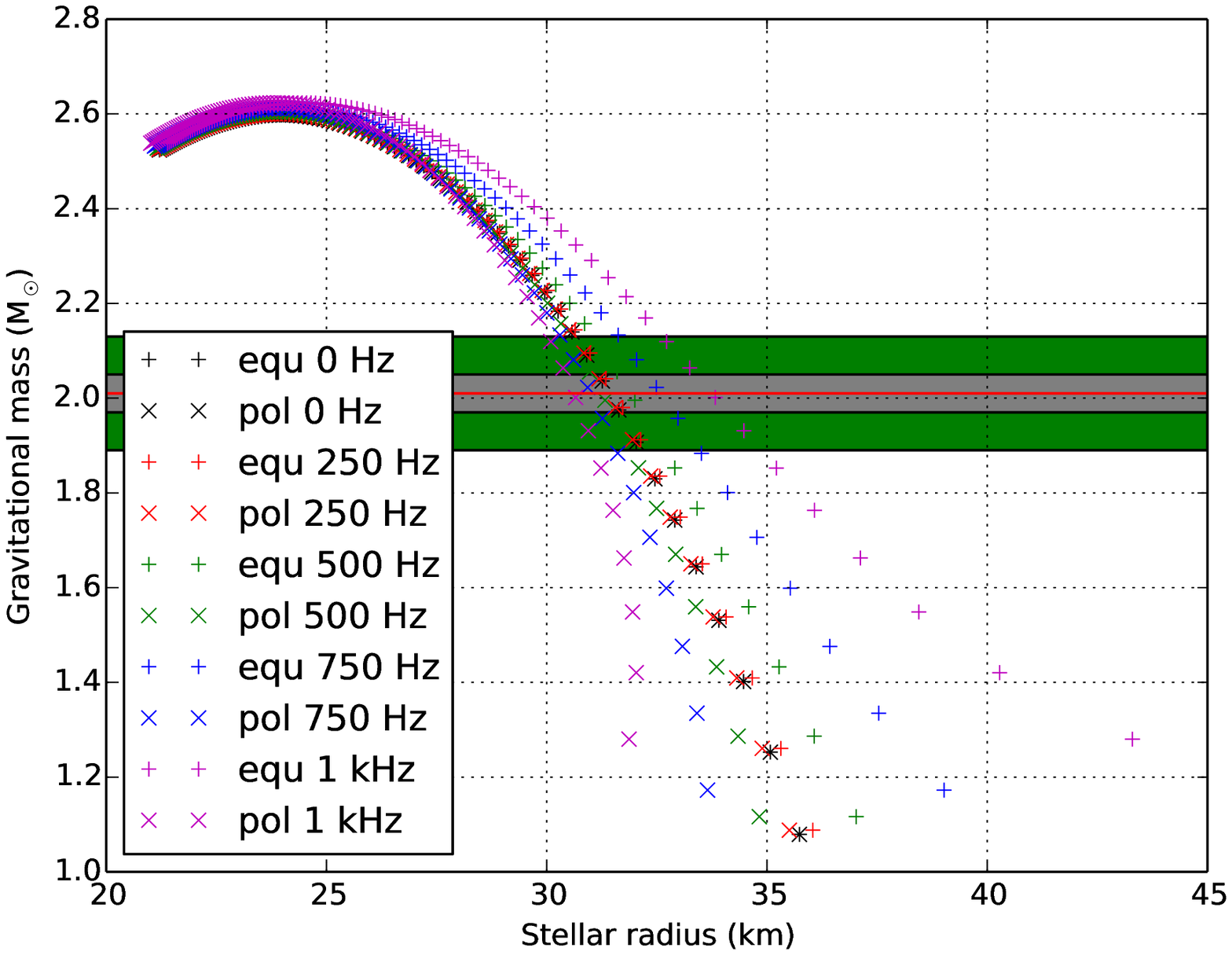} \\
\end{tabular}
\vspace*{-0.2cm}
\end{center}
\caption{The gravitational mass of neutron stars with a variety of spin frequencies and obeying the CSW EOS are plotted as a function of the central-density (left-column) and radius (right-column). Each row corresponds to a different value of the CSW EOS parameter (in cm$^{5}$ g$^{-1}$s$^{-2}$), namely, $K = 4.65 \times 10^{5}$ (top row), $K=5.58 \times 10^{5}$ (middle row), and $K=6.50 \times 10^{5}$ (bottom row). The equatorial (``equ'') and polar (``pol'') radii for a variety of neutron star spins along with the non-spinning (i.e., ``0 Hz spin'' case) are also shown here.}
\label{K-parameter_mass-radius_plot2}
\end{figure}

\begin{figure}[h]
\begin{center}
\vspace*{-0.4cm}
\begin{tabular}{c}
\hspace*{-1.25cm}
\includegraphics[width=0.6\textwidth,angle=0]{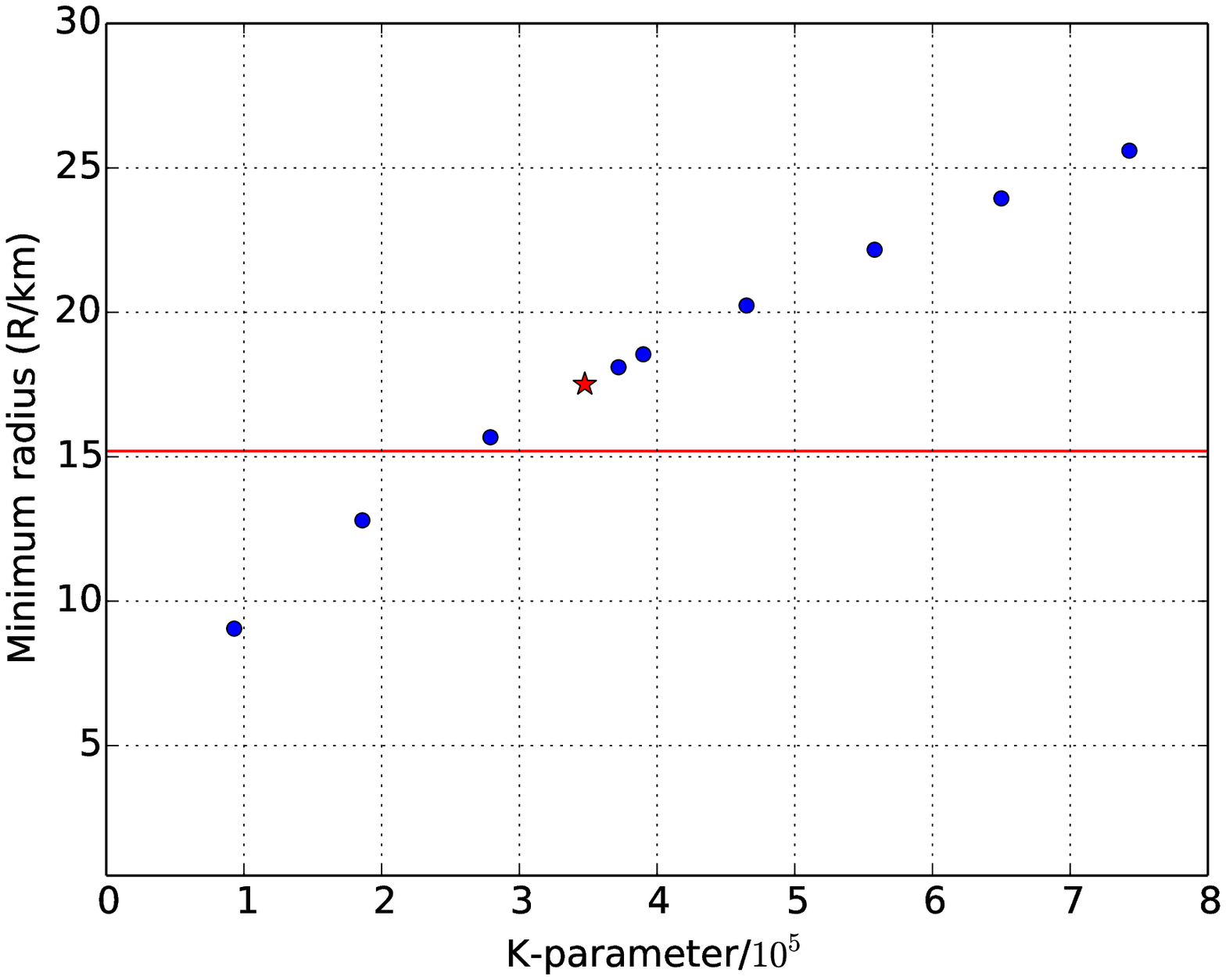} \\
\end{tabular}
\vspace*{-0.8cm}
\end{center}
\caption{The minimum radius of a non-spinning stable BEC star obeying the CSW equation of state~\cite{Colpi:1986ye} is plotted as a function of the free parameter $K$ (in cm$^{5}$ g$^{-1}$s$^{-2}$) in (blue) dots. The thin horizontal (red) line shows the observed maximum upper-bound on the radius of a neutron star~\cite{vanStraaten:2000gd}. Since the locus of dots intersects the line at $K =  2.623 \times 10^{5}$ cm$^{5}$ g$^{-1}$s$^{-2}$, this is the minimum value of $K$ that corresponds to 15.2 km radius, which is the largest among the observational value as mentioned in \ref{sec:astrobs}. The (red) star is at $K = 3.475 \times 10^{5}$ cm$^{5}$ g$^{-1}$s$^{-2}$, which is the same data point shown in Fig.\ref{K-parameter_max-mass-plot} as well.}
\label{K-parameter_min-radius-plot}
\end{figure}

Simultaneous measurements of the masses and radii of several neutron stars, as listed in the preceding section, put a very strong constraint on their equation of state. The equation of state that we considered here, namely CSW EOS, which was proposed by Chavanis and Harko \cite{Cha} as the EOS of recently observed massive neutron stars, predicts a large neutron star radius, $R > 17.5$~km, for any value of its gravitating mass, provided that the value of the free parameter $K$ is $\geq 3.475 \times 10^{5}$ cm$^{5}$ g$^{-1}$s$^{-2}$.
As the value of $K$ increases (due to increase in scattering length $a$, or decrease in Cooper pair mass $m$, or a combination of both), the value of the neutron star radius also increases to unrealistically large values, as shown in Fig.~\ref{K-parameter_min-radius-plot} for non spinning cases. The effect of spin in changing either the maximum mass or minimum radius is negligible for any value of $K$, as can be seen from Figs.~\ref{K-parameter_mass-radius_plot1} and \ref{K-parameter_mass-radius_plot2}. Thus, $K \geq 3.475 \times 10^{5}$ cm$^{5}$ g$^{-1}$s$^{-2}$ is strongly disfavored by these astronomical observations. Moreover, the existence of not-so-massive neutron stars $M \lesssim 1.6~M_\odot$ having a small radius (i.e., $R\lesssim 10$ km), as reported by G\"{u}ver et al. \cite{Guver:2010td}, implies that even for $K = 3.475 \times 10^{5}$ cm$^{5}$ g$^{-1}$s$^{-2}$ the radius predicted by the CSW EOS is too large to be consistent with those observations.

Therefore, any matter that obeys the CSW EOS is practically ruled out as the constituent of neutron stars observed in nature. Note that this conclusion 
does not preclude: (a) Boson stars obeying a different EOS, in general, or neutron stars with interior composition of BECs of kaons or pions obeying a different EOS, in particular; (b) Boson stars obeying the CSW EOS but with masses that are very different, e.g., comparable to supermassive black hole masses. In the latter case, however, one is still left with the problem of explaining why stellar mass objects are not favored while much heavier objects of the same EOS are allowed.

\section*{Acknowledgments}
This work is supported in part by NSF Grant No. PHY-1206108. AM acknowledges support from the ``SERB Start-Up Research for Young Scientists Scheme'' project grant no: SB/FTP/PS-067/2014, DST, India. We would like to thank Sharon Morsink and Nick Stergioulas for providing us with useful computational assistance. We also thank Sanjay Reddy, Michael Forbes, and Nils Anderson for helpful discussions. AM and SB thank the Institute for Nuclear Theory at the University of Washington, Seattle, for its hospitality, and the Department of Energy for partial support. AM thanks Department of Physics and Astronomy, WSU, Pullman for its hospitality. This paper has been assigned INT preprint no. INT-PUB-14-040, IUCAA Preprint no. 17/2014, and ICTS Preprint no. ICTS/2014/15.


\onecolumngrid

\onecolumngrid

\appendix
\section{A brief description of the RNS code}
\label{app:RNS}

In this paper we study the static spherically symmetric (non-spinning) as well as the stationary axisymmetric spinning equilibrium configurations of a star. In the latter case we derive configurations for four different stellar spin frequencies. We make the following three assumptions: (i) The star resides in an asymptotically flat spacetime; (ii) The spacetime has a timelike Killing vector $t^\alpha$ and an azimuthal (rotational) Killing vector $\phi^\alpha$; (iii) the two Killing vectors commute
and 
there exists a transformation, $t^\alpha \rightarrow -t^{\alpha}$ and $\phi^\alpha \rightarrow -\phi^{\alpha}$, under which the spacetime is isometric. Under these reasonable assumptions the background metric $g_{\alpha \beta}$ of a star turns out to be \cite{ButterworthIpserApJ1976,BardeenWagonerApJ1971}:
\begin{equation}
 ds^2 = - e^{2\nu} dt^2 + e^{2\psi} (d\phi - \omega dt)^2 + e^{2\mu} (dr^2 + r^2 d\theta ^2) \,,
\end{equation}
where $\nu$, $\psi$, $\omega$ and $\mu$ are metric functions that depend only on the coordinates $r$ and $\theta$. 

Furthermore, we assume the matter of the star to be a perfect fluid, which is described by the energy-momentum tensor
\begin{equation}
 T^{\alpha \beta} = (\epsilon + p) u^{\alpha} u^{\beta} + Pg^{\alpha \beta} \,,
\end{equation}
where $\epsilon$ is the energy density, $p$ is the pressure and $u^{\alpha}$ is the four-velocity of a fluid element.
The relativistic Euler equation for this stationary axisymmetric star is given by
\begin{equation}
 \frac{\bigtriangledown _{\alpha} p}{(\epsilon + p)} = -u^{\beta} \bigtriangledown _{\beta} u_{\alpha} \,.
\end{equation}
We used the publicly available RNS-code \cite{Stergioulas:1994ea} which follows the algorithm of Cook et al. \cite{Cook:1993qr}, with some minor modifications, to numerically compute the solutions of the above equation for the CSW equation of state and for stellar spin frequencies 250 Hz (shown by red curves in Figs. \ref{K-parameter_mass-radius_plot1} and \ref{K-parameter_mass-radius_plot2}), 500 Hz (green curves), 750 Hz (blue curves) and 1 kHz (pink curves) along with the non-spinning (black curves). The RNS code can compute individual stellar models as well as sequences of them for fixed mass, rest mass, angular velocity or angular momentum. The code uses uniform rotation in all models with spin, which is a very good approximation for the 
compact configurations studied here.


\end{document}